\documentclass{nature}
\usepackage{graphicx}
\makeatletter
\let\saved@includegraphics\includegraphics
\AtBeginDocument{\let\includegraphics\saved@includegraphics}
\renewenvironment*{figure}{\@float{figure}}{\end@float}
\makeatother

\usepackage{multirow}
\usepackage{subfigure}
\usepackage{mathtools}
\usepackage[colorlinks=true,linkcolor=blue,citecolor=black,urlcolor=blue]{hyperref}
\usepackage{authblk}
\usepackage[labelfont=bf]{caption}
\usepackage[figurename=Figure]{caption}
\usepackage{siunitx}
\usepackage[dvipsnames]{xcolor}
\usepackage{soul}
\usepackage{xfrac}
\usepackage{amsmath}
\usepackage{soul,xcolor}
\usepackage{ulem}

\newcommand{\samir}[1]{\textcolor{Blue}{#1}}

\newcommand{\xmrgamma}{$\text{XMR}^{\gamma}$ }

\begin{document}

\title{Spin-orbit enabled all-electrical readout of chiral spin-textures}
\author[1*]{Imara Lima Fernandes}
\author[1]{Stefan Bl\"ugel}
\author[1,2*]{Samir Lounis}
\affil{Peter Gr\"{u}nberg Institut and Institute for Advanced Simulation, Forschungszentrum J\"{u}lich and JARA, D-52425 J\"{u}lich, Germany}
\affil{Faculty of Physics, University of Duisburg-Essen and CENIDE, 47053 Duisburg, Germany}
\affil[*]{i.lima.fernandes@fz-juelich.de, s.lounis@fz-juelich.de}

\maketitle

\section*{Abstract}

\begin{abstract}
Chirality and topology are  intimately related fundamental concepts, which are heavily explored to establish spin-textures as potential magnetic bits in information technology. 
However, this ambition is inhibited since electrical reading of chiral attributes is highly non-trivial with conventional  current perpendicular-to-plane (CPP) sensing devices. Here we demonstrate from extensive first-principles simulations and multiple scattering expansion  the emergence of the chiral spin-mixing magnetoresistance (C-XMR) enabling highly efficient all-electrical readout of the chirality and helicity of respectively one- and two-dimensional magnetic states of matter. It is linear with spin-orbit coupling in contrast to the quadratic dependence associated with the newly unveiled non-local spin-mixing anisotropic MR (X-AMR). Such  transport effects are systematised on various non-collinear magnetic states -- spin-spirals and skyrmions -- and compared to the uncovered spin-orbit-independent multi-site magnetoresistances. Owing to their simple implementation in readily available reading devices, the proposed magnetoresistances offer exciting and decisive ingredients to explore with all-electrical means the rich physics of  topological and chiral magnetic objects.
\end{abstract}

\section*{Introduction}

The electrical reading of information carriers defines the viability of prospected paradigm shifts for future data storage media and logic circuits. Ideal candidates are topological magnetic textures such as individual skyrmions~\cite{Fert2013,Fert2017} due to their inherent stability ~\cite{Bogdanov1989,Roessler2006,Nagaosa2013}  and swift manipulation by currents of very low density~\cite{Jonietz2010,Fert2013,Sampaio2013,Woo2016,Jiang2016,Litzius2017} or electric field~\cite{Ma2018,Hu2018}. A successful implementation in available technologies hinges on the electrical detection of the chiral nature of these objects and on the ability to discern distinct topological spin entities, which can coexist in a given device~\cite{Koshibae2014,Leonov2015,Ritzmann2018,Rybakov2019,Yokouchi2020,Ritzmann2020,jena2020,Gao2020,Heigl2021,Yao2020}. Over the last years, control of spin-chirality, settled by the direction of the Dzyaloshinskii-Moriya interaction (DMI)~\cite{Dzyaloshinsky1958,Moriya1960}, was demonstrated experimentally. For instance, reversible chirality transition was realized via hydrogen chemisorption and desorption~\cite{Chen2021} while the application of an  electric field or ultrashort electromagnetic pulse  induce large modifications of DMI~\cite{Srivastava2018,Buettner2021} in accordance with theoretical predictions~\cite{Chshiev2018,Yao2020,Desplat2020}.

\begin{figure}[h!]
\centering
\includegraphics[width=0.9\columnwidth,keepaspectratio]{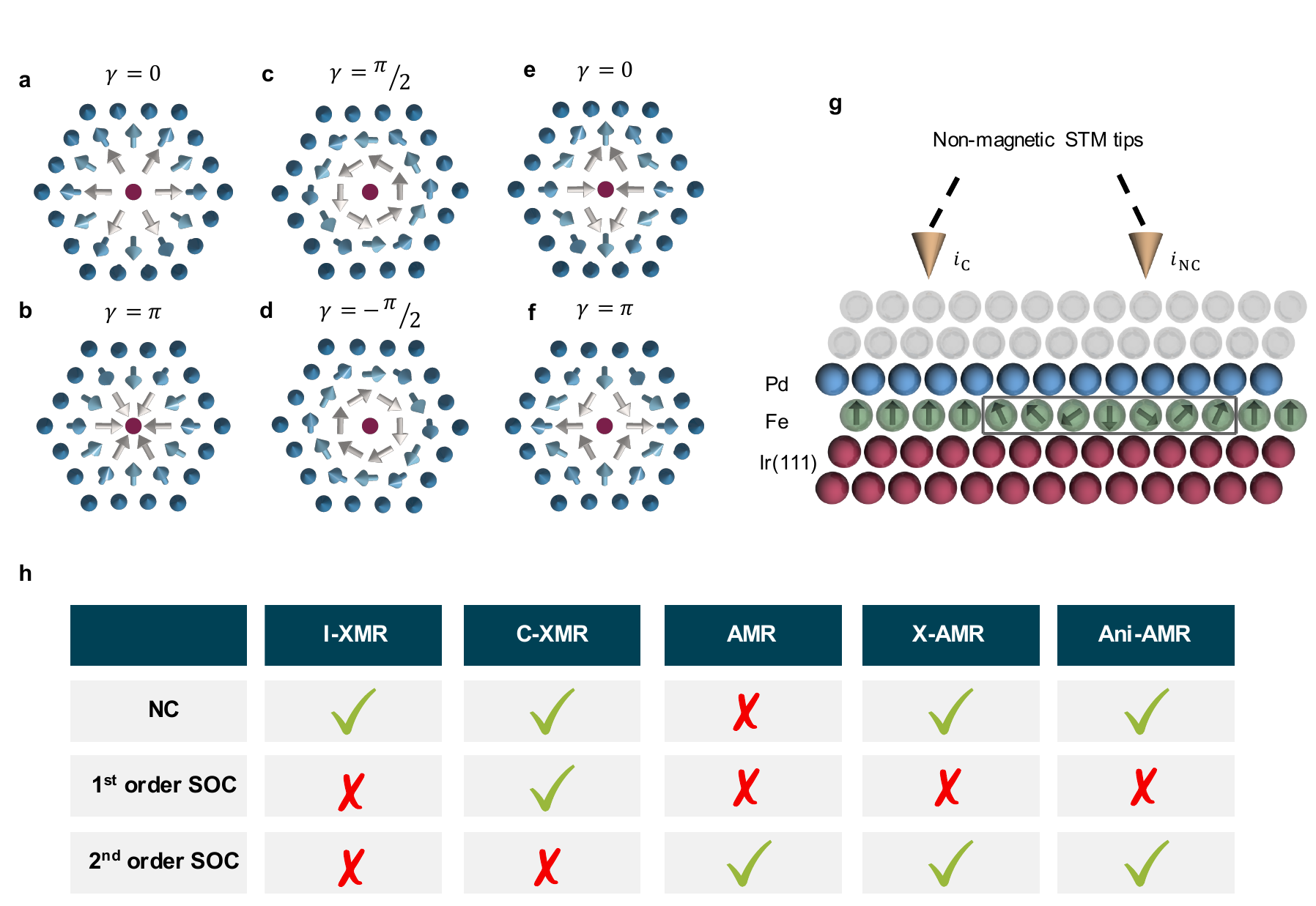}
\caption{\textbf{ Examples of various magnetic skyrmions.}  Néel-type skyrmion ($Q=-1$) exhibits a \textbf{a} counter-clockwise cycloidal spin rotation, e.g, $\gamma=0$ or a \textbf{b}  clockwise   cycloidal spin rotation,  e.g, $\gamma=\pi$. The Bloch skyrmion exhibits a helical spin rotation with helicity a \textbf{c} $\gamma=\pi/2$ or \textbf{d} $\gamma=-\pi/2$. The antiskyrmion ($Q= 1$) with helicity \textbf{e} $\gamma=0$ and \textbf{f} $\gamma=\pi$. \textbf{g} Illustrative STS experiment detecting the electronic signals emanating from either a magnetic skyrmion or the ferromagnetic background. The sample consists of fcc-Pd (blue spheres) atop an Fe overlayer (green spheres) on single-crystal fcc-Ir(111) bulk substrate (red spheres). The tip probes the local density of states (LDOS) two vacuum layers (grey spheres) away from Pd. \textbf{h} 
Classification of the different unveiled magnetoresistances as function of triggering mechanisms: non-collinear (NC) magnetism, 1$^\text{st}$ and 2$^\text{nd}$ order SOC.
}
\label{fig:Fig1}
\end{figure}

Identifying the vector spin-chirality, which is defined by the cross-product of adjacent magnetic moments is pertinent when addressing one-dimensional magnetic states such as spin-spirals and domain walls. This notion is, however, complemented with concepts such as  helicity $\gamma$ and topological charge $Q$ when dealing with two-dimensional spin-textures~\cite{Nagaosa2013} (see Figure.~\ref{fig:Fig1}a-f and Supplementary Note 1). $Q$ measures the wrapping of the magnetization around a unit sphere and enables to discern skyrmions ($Q=\samir{-}1$) from antiskyrmions ($Q=\samir{1}$) or magnetic vortices ($Q=0.5$). $\gamma$ is defined by the in‐plane magnetic moments swirling direction (e.g., clockwise or counterclockwise) and distinguishes magnetic textures of identical topological charge such as Bloch- from N\'eel-type skyrmions. While  N\'eel-type skyrmions exhibit a right-handed or left-handed cycloidal spin rotation leading to a helicity of $\gamma=0$ or $\gamma=\pi$, respectively, Bloch-type skyrmions host helical spin rotation with a helicity of $\gamma=\pm \pi/2$.

 Electric-readout of chirality  has rarely been possible with conventional technologies based on current perpendicular-to-plane (CPP) geometries.   Resting on tunnel magnetoresistance (TMR), which requests two magnetic electrodes, spin-polarized scanning tunneling microscopy/spectroscopy (STM/STS) was so far the only  technique successful in detecting the chirality of spin spirals and sub-10 nm skyrmions~\cite{Heinze2011,Romming2013,Romming2015,Meyer2019,Perini2019,Muckel2021}. The challenge resides in the highly non-trivial full control of the magnetization direction of the readout electrode and the sensing efficiency, dictated by the  relative alignment of the magnetization of two electrodes. Obviously, the realization of  all-electrical chiral detection, i.e.\ readout with a non-magnetic electrode, would enable a remarkable leap forward on the basis of a robust  practical implementation. 
However, the known MR effects such as the anisotropic MR (AMR)~\cite{McGuire1975,Bode2002,Gould2004,Matos-Abiague2009} and the recently discovered highly-efficient spin-mixing MR (XMR)~\cite{Crum2015,Hanneken2015} are expected to  be transparent to the main attributes of topological structures: chirality or helicity. With an efficiency of a few percent, the AMR originates from spin-orbit coupling (SOC) and is by nature an on-site effect that depends on the local quantization axis of the magnetization relative to the crystal lattice. The X in XMR stands for spin-mixing since the XMR effect has been proposed to  mainly arise from the spin-mixing of the electronic states as a result of a non-local mechanism driven by electron scattering at atoms having canted magnetic moments independently from SOC~\cite{Crum2015,Hanneken2015}.  Intriguingly, the precise nature of the dependence of measured signals on the opening angles of the underlying spin-textures remained so far  elusive~\cite{Kubetzka2017}.

In contrast to CPP geometries, all-electrical chiral and topological characterization is in principle possible with  current-in-plane (CIP) transport concepts such as the topological Hall effect~\cite{Bruno2004} or the recently proposed chiral~\cite{Lux2020} and non-collinear~\cite{Bouaziz2021} Hall effects.  However, these effects are fairly small in general and have to be disentangled from a large Hall background. Moreover, the required Hall bar junction complicates their integration in standard device structures~\cite{Neubauer2009,Lee2009,Zhang2016,Maccariello2018,Zeissler2018}.

In this work, we unveil the existence of a rich family of MR effects enabling the all-electrical perpendicular readout of distinct chirality or helicity characteristics of spin-textures. We utilize multiple scattering concepts and intensive systematic first-principles simulations of atomically resolved transport measurements as probed within STM/STS on   magnetic textures generated in a prototypical material: fcc-PdFe bilayer  deposited on Ir(111) surface (see Methods). The chosen substrate is well known~\cite{Romming2013,Crum2015,Romming2015,Dupe2014,Simon2014,Dias2016} to host isolated Néel-type skyrmions with a few nanometers diameter.

The newly discovered set of MRs are categorized in terms of the presence (or not) of  non-collinear magnetism, SOC and their concomitant intertwining, as schematically illustrated in Fig.~\ref{fig:Fig1}h.  Purely non-collinear magnetism, i.e. without SOC, gives rise to an isotropic MR induced by the spin-mixing of electronic states due to the misalignment of the atomic magnetic moments, which is the transport phenomenon usually discussed in Refs.~\cite{Crum2015,Hanneken2015}.  We name it here I-XMR since it is  isotropic and transparent to SOC related phenomena, as shown from our analytical derivations and  systematic ab-initio simulations. 
We identify the novel chiral XMR effect (C-XMR), which is linear in SOC and requires non-collinear magnetism as well as broken inversion symmetry similarly to the conditions giving rise to DMI.  C-XMR reaches large efficiencies permitting to distinguish and spatially probe the chiral nature of a given magnetic object. 
Traditional AMR is quadratic in SOC and emerges in a collinear magnetic environment upon a global rotation of the spin moments. In our study, however, we found a spin-mixing AMR (X-AMR) effect, which is also quadratic in SOC but it is non-local and depends on the canting angle between magnetic moments. In contrast to AMR, its magnitude can be as large as conventional MR efficiencies. The peculiar angular and spatial dependencies characterizing each of the different MR mechanisms enable their straightforward experimental detection and project a plethora of opportunities for basic and applied research of chiral and topological magnets.

\section*{Results}

\subsection{Impact of helicity on the XMR signal.}

According to the Tersoff-Hamann model~\cite{Tersoff1983},  the local density of states (LDOS) decaying from a particular site $i$ into the vacuum is proportional to the differential conductance $dI/dV$ at a bias voltage $V_{\text{bias}}$ relative to the highest occupied state of the sample measured with STM employing a non-magnetic tip. The latter is assumed in the second vacuum layer above the Pd substrate, as shown in Fig.~\ref{fig:Fig1}g. The efficiency of the XMR effect is calculated as the ratio between the deviation of the conductance on top of a skyrmion from that of a reference point, which we choose to be the collinear magnetic region (see Methods section):

\begin{equation}
    \text{XMR}^\gamma(E) = \frac{\text{LDOS}^\gamma_\text{NC}(E) - \text{LDOS}_\text{C}(E)}{\text{LDOS}_\text{C}(E)}\,,
    \label{eq:xmr}
\end{equation}
where C and NC correspond to the collinear and non-collinear magnetic areas respectively, as shown in Figure \ref{fig:Fig1}, while $\gamma$ indicates the helicity of the skyrmion. 

\begin{figure*}
\centering
\includegraphics[width=\columnwidth,keepaspectratio]{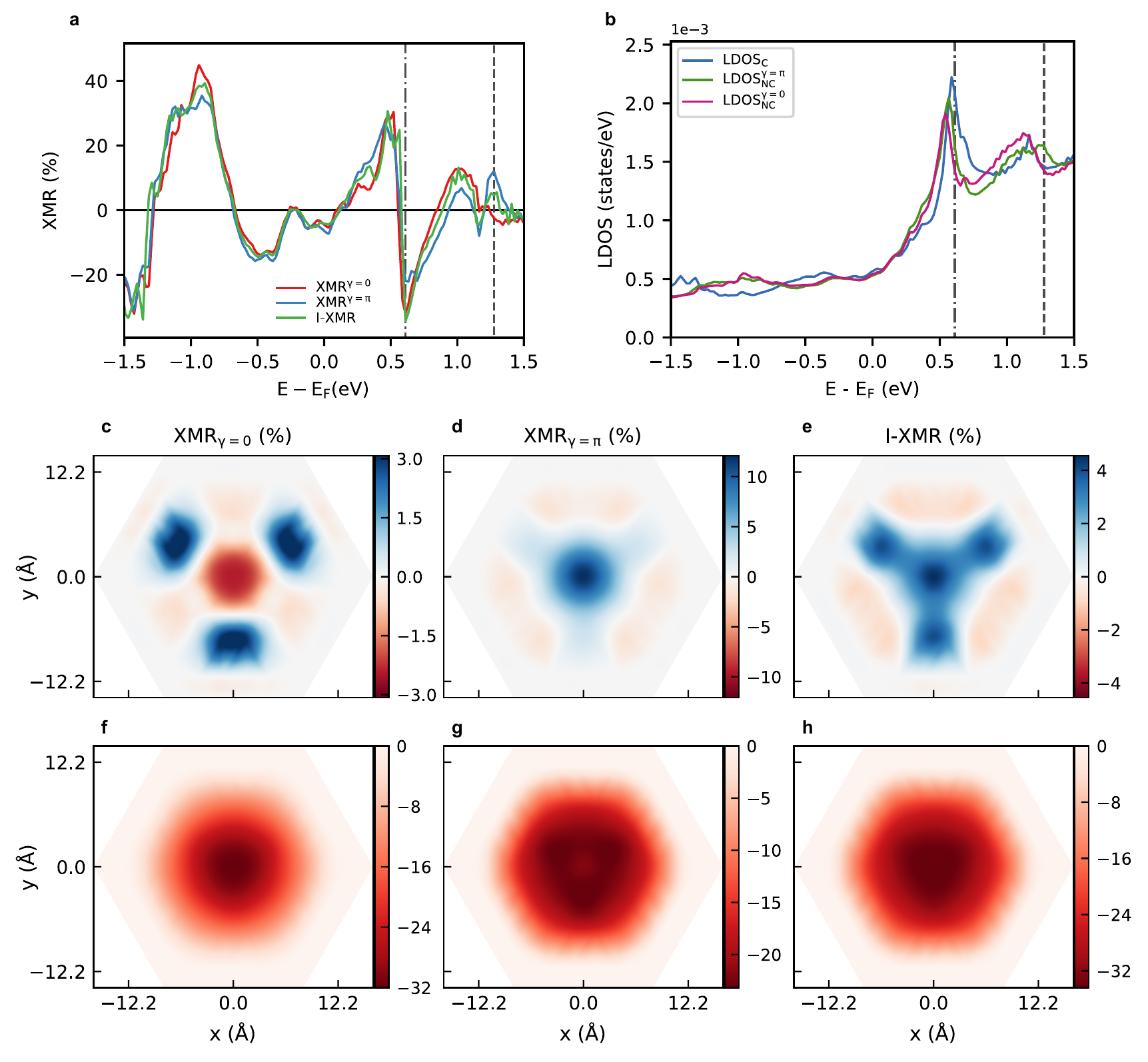}
\caption{\textbf{Impact of the helicity on the XMR signal.}  \textbf{a} Comparison of energy-resolved XMR signals with different helicities and without SOC  measured on top of the skyrmion’s core. \textbf{b} Sum of spin-resolved  electronic structure in the vacuum on top of the core of a Neéel-type skyrmion with diameter $D_{\text{sk}} \approx \SI{2.2}{\nm}$ for different helicities and above the collinear configuration. The dashed and doted-dahed  lines indicate, respectively, the tunneling energies $eV_{\text{bias}} = \SI{1.27}{\electronvolt}$ and $\SI{0.61}{\electronvolt}$ at which a spatial map of the whole skyrmion is plotted for \textbf{c, g} $\text{XMR}^{\gamma=0}$,  \textbf{e, h} $\text{XMR}^{\gamma=\pi}$ and  \textbf{f, i} \text{I-XMR} signals.}
\label{fig:xmr_helicity}
\end{figure*}

As understood currently, the XMR effect results mainly from the non-collinearity of the magnetic moments without invoking spin-orbit coupling.  The latter, however, can affect the measured XMR signal due to AMR. As aforementioned, both effects are chiral- or helicity-independent. This implies that XMR should be unaltered when switching  the sense of rotation of the magnetic moments. 
This particular scenario is investigated by comparing the energy-resolved   \xmrgamma ratios  to the I-XMR one obtained by switching-off SOC, as illustrated in Figure~\ref{fig:xmr_helicity}a.  For a systematic comparison, the relative angles are maintained fixed to that obtained from full self-consistency of the skyrmion with $\gamma = 0$. The sense of rotation of moments has been switched to realize the skyrmion with $\gamma = \pi$ (see Figure~\ref{fig:Fig1}a-b). The spectra were obtained on top of the core of a N\'eel-type skyrmion with diameter $D_{\text{sk}} \approx \SI{2.2}{\nm}$. 
While at first sight the $\text{XMR}^{\gamma=0}$ and $\text{XMR}^{\gamma=\pi}$ signals seem to have a similar shape, they remarkably do not lay on top of each other.
Taking as an example the tunneling energy of $\SI{1.27}{\electronvolt}$ (see dashed line in Figure~\ref{fig:xmr_helicity}a), we find that SOC enhances $\text{XMR}^{\gamma=\pi}$ by a factor of three in strong contrast to the drastic reduction and even change of sign imposed on  $\text{XMR}^{\gamma=0}$. Such discrepancy is not limited to the core of the skyrmion but is a general effect, as demonstrated in the spatial maps plotted in Figure~\ref{fig:xmr_helicity}c-e. The three maps are indeed very different, with the signal collected on $\text{XMR}^{\gamma=\pi}$ being the largest. At an energy of $\SI{0.61}{\electronvolt}$ (dot-dashed line in Figure~\ref{fig:xmr_helicity}a), SOC seems now to reduce the XMR emanating from both skyrmions  but favoring the observation of the skyrmion with helicity $\gamma =0$ instead of the one with $\gamma =\pi$. Impressively,  the XMR signal without SOC peaks up at a value of $\approx35$\%, but with SOC the XMR efficiency crashes down  to $\approx24$\% for $\gamma=\pi$. The patterns found for $\text{XMR}^{\gamma=0,\pi}$ cannot be explained by the %\imara{previously} 
{hitherto} known mechanisms: merely SOC-independent non-collinear magnetism or chiral-insensitive AMR. Overall, this evidences the existence of new MR effects at play. 

The alteration of the XMR signals results certainly  from the change in the electronic structure imposed by  the reversal of the skyrmion's helicity as illustrated in Figure~\ref{fig:xmr_helicity}b, where a comparison is displayed between the LDOS measured on top of the skyrmionic (core) and collinear configurations. 
For the particular bias voltages addressed earlier, SOC reduces the predicted $dI/dV$ amplitude  at $\SI{0.61}{\electronvolt}$ and enhances at $\SI{1.27}{\electronvolt}$ the signal pertaining to the skyrmion with $\gamma = \pi$ without affecting that of the zero-helicity skyrmion. 
In general, the shape, linewidth and energy of the electronic states depend simultaneously on SOC and on the chirality of the underlying magnetic configuration. Clearly, SOC enables electronic  hybridization channels  sensitive to the sense of rotation of the magnetic moments.

\subsection{Multi-site isotropic, chiral and anisotropic XMR signals.}

A deeper understanding of the unveiled observations can be unravelled by scrutinizing the different contributions  to the LDOS, which are  enabled by the rotation of magnetic moments. In the process of building up the electronic structure at a given site, electrons scatter in the lattice and hope between neighboring magnetic atoms, picking up various information.  Using multiple-scattering theory (for details see Supplementary Information 2), we demonstrate that local and non-local multi-site contributions of different nature emerge: isotropic (independent from SOC), chiral (linearly dependent on SOC) and anisotropic achiral (quadratically dependent on SOC), which are strongly contingent to the underlying magnetic texture. 

We find that the all-electrical XMR signal at a site $i$ is a sum of four terms:
\begin{eqnarray}
    \text{XMR(E)}= \text{I-XMR}(E) + \text{C-XMR}(E) + \text{X-AMR}(E) + \text{Ani-XMR}(E)\, . 
    \label{Eq_XMR}
\end{eqnarray}

We start with the most basic contribution, $\text{I-XMR}$, which is isotropic and does not require SOC. As shown in Supplementary Note 2, it primarily arises from a two-site isotropic term, ${\gamma}_\text{I-XMR}^\text{2-spin}(E)\; \mathbf{S}_i\cdot\mathbf{S}_j$, which quantifies the change of the LDOS of atom $i$ because of  the misalignment of its moment from the one carried by a neighboring atom $j$ independently from SOC. Although the coupling can in principle be long-ranged, we limit our discussion here to nearest neighbors. This indicates that the magnetoresistance %, denoted in this particular case as $\Delta \text{LDOS}_\text{NC}(E)$,
should change in a cosine fashion upon rotation of the moments, with a maximum expected when the moments are aligned antiferromagnetically. In the adiabatic approximation, i.e.\ when the skyrmions are large, it is then expected that the isotropic contribution should be quadratic with the rotation angle between two magnetic moments.
Similarly to the extended Heisenberg Hamiltonian, one can identify an additional biquadratic contribution $\gamma_\text{I-XMR}^\text{4-spin}(E)(\mathbf{S}_i\cdot\mathbf{S}_j)^2$ and even other higher-order terms.

Besides the isotropic contributions we uncover a chiral term, $\text{C-XMR}$, linear with SOC and directly proportional to the chiral product between the spin-moment $i$ and its  neighboring moments:  ${\gamma}_\text{CXMR}(E)\;\mathbf{\hat{c}}_{ij}\cdot\left(\mathbf{S}_i\times\mathbf{S}_j\right)$, where $\mathbf{\hat{c}}$ is analogous to the DM vector and obeys therefore similar symmetry rules.  
Owing to the chiral nature of this term, rotating the moments clockwise or counter-clockwise modifies the LDOS in an opposite fashion. The largest contribution arises if the rotation angle is $\pi/2$. For large skyrmions, the angular dependence is linear with the angle between moments. Obviously, {since the remaining contributing terms are achiral}, by switching the helicity of the skyrmion (or the chirality of a spin-spiral or a domain wall), one can immediately extract the chiral XMR:
\begin{eqnarray}
\text{C-XMR}^{\gamma =0}(E) = - \text{C-XMR}^{\gamma =\pi}(E)= \frac{1}{2}\left(\text{XMR}^{\gamma =0}(E) - \text{XMR}^{\gamma =\pi}(E)\right).
\end{eqnarray}

The usual AMR is quadratic in SOC and is local by nature since it {only} depends on the orientation of moment $i$~\cite{McGuire1975,Bode2002,Gould2004,Matos-Abiague2009}. Conventionally, it is written as $\gamma_\text{AMR}(E) (\mathbf{e^z \cdot \mathbf{S}_i})^2$, assuming that the anisotropy field responsible for the AMR effect points along the cartesian direction $z$. 
In our multiple scattering expansion, however, we uncover non-local (multi-site) dependencies generalizing the  AMR to what we dub spin-mixing AMR (X-AMR) effect. The latter  is proportional to $\gamma_\text{X-AMR}(E) (\mathbf{e^z \cdot \mathbf{S}_i})(\mathbf{e^z \cdot \mathbf{S}_j})$. So here the angular dependence is $\cos{\theta_i}\cos{\theta_j}$ instead of the expected local angular dependence $\cos^2{\theta_i}$, where $\theta_i$ is the polar angle pertaining to moment $i$.  Furthermore, we find a SOC-correction to the isotropic two-site term: ${\gamma}_{\text{Ani-XMR}} (E)\, \mathbf{S}_i \cdot \mathbf{S}_j$.

Owing to the distinct angular dependence of each of the identified MRs, it is possible to fit the spatial maps and extract the magnetoresistance weights associated to each mechanism. We can expand Equation~\ref{Eq_XMR}: $\text{XMR}(E)=\sum_{i\neq j} \text{XMR}_{ij}(E)$ with the sum limited to the nearest neighbors and :
\begin{eqnarray}
    \text{XMR}_{ij}&=& \left[\gamma_\text{I-XMR}^\text{2-spin}+ \gamma_{\text{Ani-XMR}}\right]  \mathbf{S}_i\cdot\mathbf{S}_j +
    \gamma_\text{I-XMR}^\text{4-spin}(\mathbf{S}_i\cdot\mathbf{S}_j)^2+\nonumber\\
    &&
    {\gamma}_\text{C-XMR}\;\mathbf{\hat{c}}_{ij}\cdot\left(\mathbf{S}_i\times\mathbf{S}_j\right) 
    + \gamma_\text{X-AMR} (\mathbf{e^z \cdot \mathbf{S}_i})(\mathbf{e^z \cdot \mathbf{S}_i}) \;,
\end{eqnarray}
where the energy dependence is omitted for simplicity.

We expect the spatial map of each contribution to show distinctive patterns depending on the underlying spin-texture. The overlap of all terms can be complex and hinges on the MR weights magnitude and sign. In Table~\ref{tab:parameters}, we list the weight  distributions for three bias energies:  0.56, 0.61 and $\SI{1.27}{\electronvolt}$ as obtained for the skyrmion studied in Figure~\ref{fig:xmr_helicity} (see Supplementary Note 3-5 for further details). One immediately remarks that all contributions can be of the same order of magnitude, with their sign and size being energy-dependent. We note that in contrast to the AMR weight, the contribution of the rest of the parameters will be enhanced due to the non-local nature of their associated MRs. On the fcc-111 surface, the maximum  enhancement factor of 6 is expected due to the number of nearest neighboring pairs. Remarkably, the weight pertaining to X-AMR is larger than that of the conventional AMR. Even if quadratic with SOC, it can  directly compete with the weight characterizing C-XMR, which is linear with SOC.  The SOC-independent XMR seems to provide  the largest contribution to the electrical signal, however, the newly unveiled SOC-driven MRs are significant. We note that at  $\SI{1.27}{\electronvolt}$, the signal is dominated by C-XMR with a weight of opposite sign to the conventional I-XMR.

\begin{table*}
\footnotesize
\centering
\caption{\textbf{Energy-dependent weights of the various unveiled spin-mixing magnetoresistances}. The tunneling energies are given in eV units.}
\begin{tabular}{ c c c c c c c}
\hline 
\hline \noalign{\smallskip}
Energy & $\gamma_{\text{I-XMR}}^{{\text{2-spin}}}$ &  $\gamma_{\text{I-XMR}}^{{\text{4-spin}}}$ &  $\gamma_{\text{C-XMR}}$&  $\gamma_{\text{AMR}}$ &  $\gamma_{\text{X-AMR}}$ &  $\gamma_{\text{Ani-XMR}}$ \\
\noalign{\smallskip}\hline\noalign{\smallskip}
0.56 & $-$6.4190  &  0.0361 &   3.7392 &  -0.9416  & 4.7319  & 0.0107 \\ \noalign{\smallskip}\hline\noalign{\smallskip}
0.61 & 12.4692   &     -0.1588   &      2.0918   &  0.7079   &  -3.3460         &     0.1750 \\ \noalign{\smallskip}\hline\noalign{\smallskip}
1.27 & -1.0456    &     0.0045     &    3.1518 &  0.0206  &  0.0442        &        -0.0443 \\ \noalign{\smallskip}\hline\noalign{\smallskip}
\hline
\end{tabular}
\label{tab:parameters}
\end{table*}

\begin{figure*}
\centering
\includegraphics[width=1.0\columnwidth,keepaspectratio]{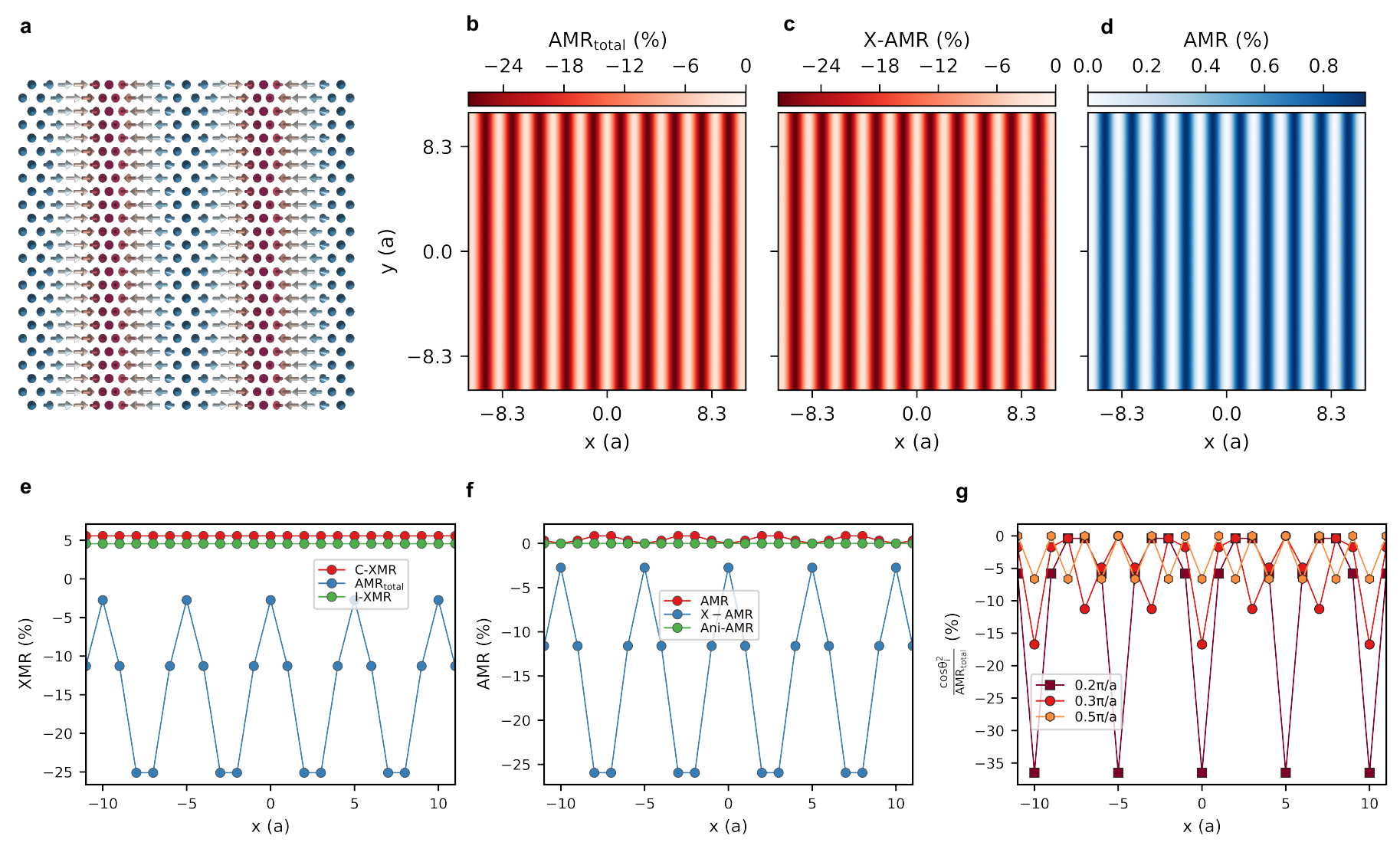}
\caption{\textbf{ Magnetoresistance patterns characterizing spin-spirals.}  \textbf{a} Spin-spiral with wave vector $q = 0.2\pi/\text{a}$ along the [100] direction, with $a$ being the inplane lattice constant of Ir(111), i.e., $a = a_\text{bulk}/\sqrt{2}$ where $a_\text{bulk} = \SI{3.83}{\angstrom}$, and associated AMR signals: \textbf{b} total AMR, \textbf{c} X-AMR and \textbf{d} conventional AMR considering the MR weights obtained at a bias energy of  $eV_{\text{bias}} = \SI{0.56}{\electronvolt}$. A systematic comparison of the different XMR components along the $x$-direction is provided in \textbf{e} and \textbf{f}. The ratio $\cos^2{\theta}_i / \text{AMR}_\text{total}$ is shown in \textbf{g} for spirals of wave vectors $0.2\pi/\text{a}$, $0.3\pi/\text{a}$ and $0.5\pi/\text{a}$ to visualize the contribution of the new X-AMR.}
\label{fig:spin-spiral}
\end{figure*}

Before proceeding to the skyrmion case, it is instructive to explore the behavior of the different MRs for the case of a homogeneous spin-spiral (see Figure~\ref{fig:spin-spiral}) utilizing MR weights provided in Table~\ref{tab:parameters} at an energy bias of  $eV_{\text{bias}} = \SI{0.56}{\electronvolt}$. We note that the PdFe bilayer on Ir(111) surface hosts spin spirals  when the magnetic field is off. Since the rotation angle of the magnetic moments is constant, I-XMR, Ani-XMR and C-XMR are not altered across the whole homogeneous spirals and can only be probed  upon modification of the spin-spiral pitch (or size of a domain wall). If one manages to switch the chirality of the spiral with external means~\cite{Chshiev2018,Yao2020,Srivastava2018,Desplat2020,Buettner2021,Chen2021}, C-XMR would change sign. 
In general, the spin-spiral patterns of the probed all-electrical MR emerge uniquely from both the AMR and X-AMR, summing up to the total AMR (Figure~\ref{fig:spin-spiral}e). At first sight, they look all similar (Figure~\ref{fig:spin-spiral}b-d), but a closer inspection of their angular dependence reveals differences owing to their local versus non-local nature (Figure~\ref{fig:spin-spiral}f). By dividing the conventional $\cos^2(\theta)$ characterizing the traditional AMR by the total AMR, one can identify the signature of X-AMR as exemplified in Figure~\ref{fig:spin-spiral}g for different spiral pitches. Increasing the wave vector of the spiral enlarges the rotation angle of the magnetic moments, which enhances (diminishes) the contribution of the X-AMR (conventional AMR).  The latter  provides a well-defined path for the experimental detection of X-AMR.

\begin{figure*}
\centering
\includegraphics[width=\columnwidth,keepaspectratio]{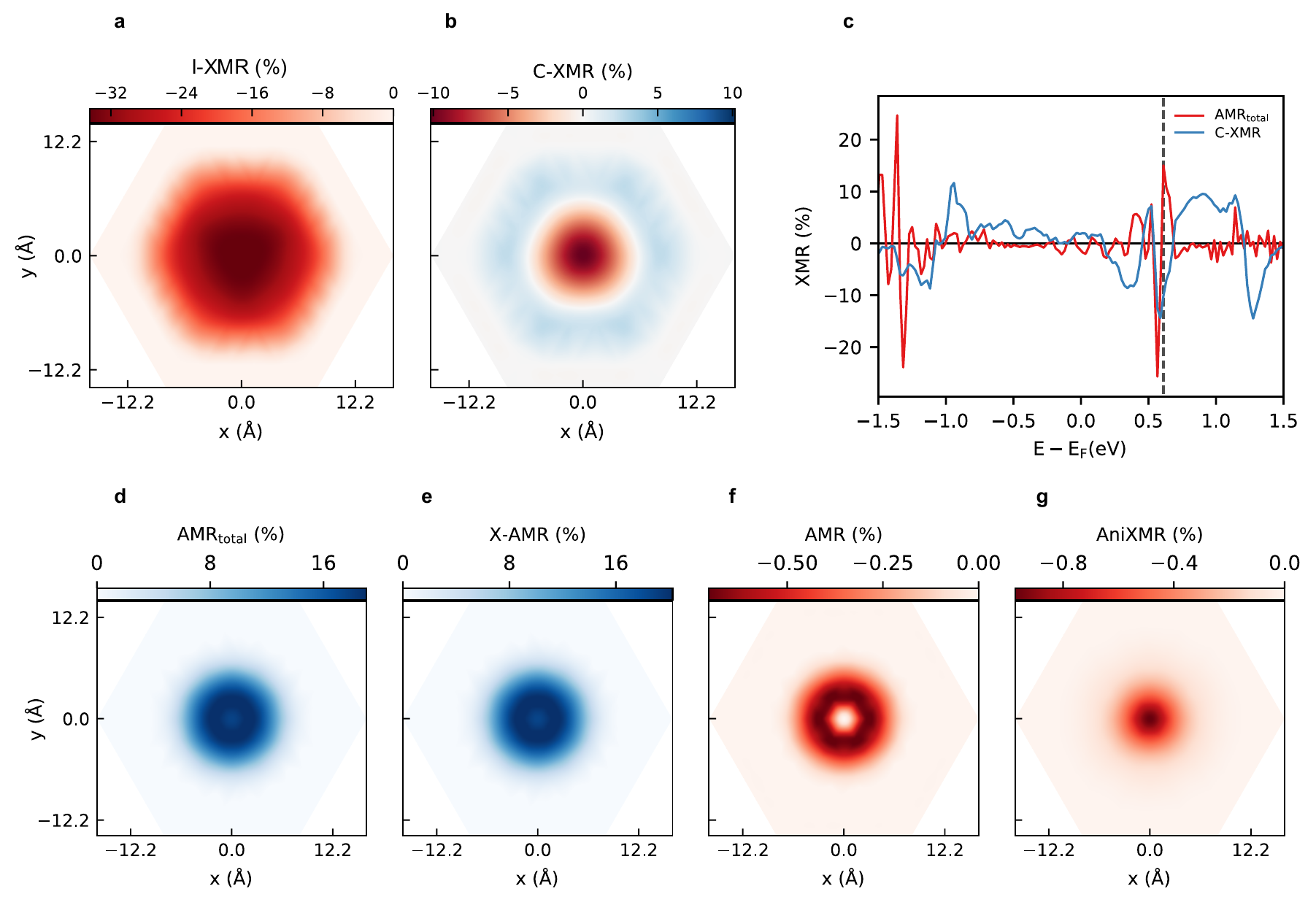}
\caption{\textbf{ Magnetoresistance patterns characterizing a N\'eel skyrmion.}  \textbf{a-b} Spin-orbit independent XMR pattern compared to the one emerging from C-XMR as obtained at the energy bias of $eV_{\text{bias}} = \SI{0.61}{\electronvolt}$. \textbf{c} Comparison of the energy-resolved C-XMR and total AMR efficiencies measured atop the core of the skyrmion. The dashed line indicates the energy at which the two-dimensional maps are extracted. 
\textbf{d} Total AMR signal decomposed into the  \textbf{e} X-AMR, \textbf{f} conventional AMR, and \textbf{g} ani-XMR components.} \label{fig:amr_c-xmr}
\end{figure*}

In contrast to spin-spirals, skyrmions are inhomogenous spin-textures that trigger spatial modulations of all XMR components. This is  visualized in Figure~\ref{fig:amr_c-xmr} for the various MR patterns hosted by a N\'eel skyrmion at $\SI{0.61}{\electronvolt}$ (see Supplementary Note 6 for the MR patters of different skyrmionic structures). The dissimilarity of the  shape, extension and sign of the plotted efficiencies is remarkable. On the one hand, the C-XMR and I-XMR culminate to their maximum value (respectively -10.2\% and -34.4\%) at the skyrmion core (Figure~\ref{fig:amr_c-xmr}a-b). On the other hand, C-XMR experiences a change of sign  close to  the skyrmion's edge, contrary to I-XMR. Interestingly, the total AMR contrast is positive and narrower than the negative ones obtained for I-XMR and C-XMR (Figure~\ref{fig:amr_c-xmr}d). Keeping in mind that the sign of C-XMR switches with the change of helicity, this is a nice example demonstrating how different MRs can counter-act each other when reading chiral magnetic objects. The total AMR signal exhibits a maximum of 19\% slightly off-centered from the skyrmion's core. At the latter location, the total AMR value drops down to 17.6\%, which is expected to be zero, i.e., similar to that of the ferromagnetic background, when counting only on the traditional local AMR.   
In this particular case, the non-local contribution X-AMR, which is two orders of magnitude larger than the conventional AMR or the Ani-XMR (Figure~\ref{fig:amr_c-xmr}e-g), dominates the total AMR efficiency. While at $\SI{0.61}{\electronvolt}$ the SOC-driven signals, i.e., C-XMR and total AMR, are equally important, albeit different sign, we identify in Figure~\ref{fig:amr_c-xmr}c the energy window ranging from $0.7$ to $\SI{1.4}{\electronvolt}$ where the MR is mostly of chiral origin. At about $-\SI{1.36}{\electronvolt}$, the total AMR  dominates the scene by reaching an impressive efficiency of 24.7\%.

\subsection{Other chiral spin textures.}   
\begin{figure*}
\centering
\includegraphics[width=\columnwidth,keepaspectratio]{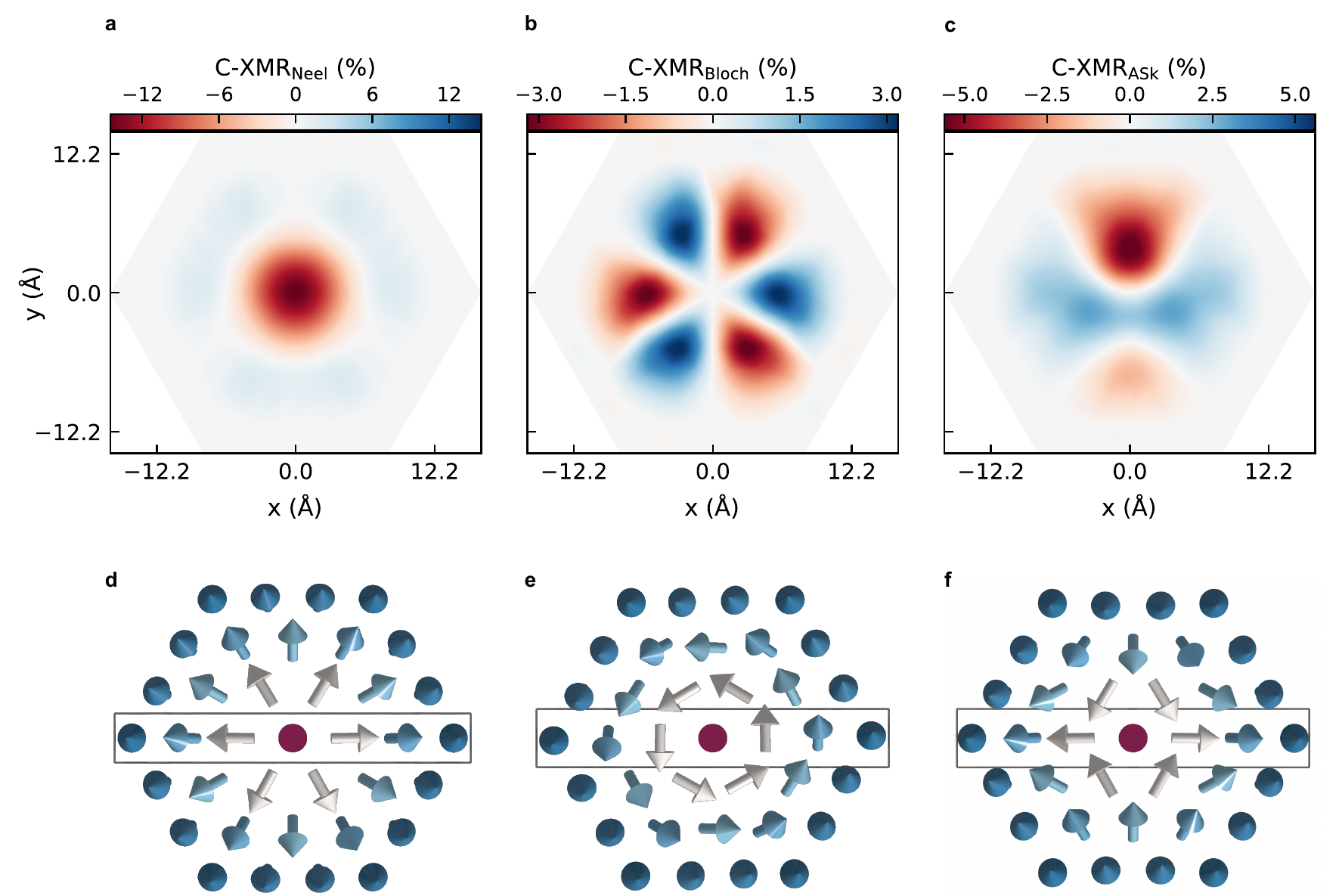}

\caption{ \textbf{C-XMR signals on various chiral skyrmions.} On the upper panel the C-XMR signal at a bias energy of $eV_{\text{bias}} = \SI{1.27}{\electronvolt}$ is shown for \textbf{a} N\'eel type and \textbf{b} Bloch type skyrmions and for \textbf{c} anti-skyrmions. On the lower panel  the corresponding skyrmion magnetic textures with diameter $D_{\text{sk}} \approx \SI{2.2}{\nm}$ are shown. The rectangular boxes are used in the text to address the change of chirality across the different skyrmions.} \label{fig:other_structures}
\end{figure*}

For completeness, we have expanded our investigations to other skyrmionic structures such as Bloch skyrmions and antiskyrmions (Figure~\ref{fig:other_structures}d-f). Henceforth, we focus on the  C-XMR effect, which is a primary tool  to distinguish chiral spin textures. We assume the anisotropy fields associated with the N\'eel skyrmion of zero helicity and plot in Figure~\ref{fig:other_structures}a-c the spatial C-XMR patterns at $\SI{1.27}{\electronvolt}$.  On the one hand, the C-XMR of the N\'eel-type skyrmion is isotropic, reaches a maximum value at its core, and reduces in magnitude when approaching the edge. On the other hand, the C-XMR patterns obtained for the Bloch-type skyrmions and antiskyrmions are strongly  anisotropic with off-centered maxima,  reflecting their chiral texture as  described by $\mathbf{\hat{c}}_{ij}\cdot(\mathbf{S}_i\times\mathbf{S}_j)$.  Here regions of zero C-XMR arise within the skyrmionic area contrary to what is observed in N\'eel skyrmions. 
In the Bloch skyrmion, one finds a sequence of six large-amplitude lobs separated by zero-lines whenever the chiral signal changes sign. In this particular case, 
the dot product of spin-chirality vector and the underlying DMI vector switches within each row of atoms when passing across the skyrmion's core (see the box shown in Figure~\ref{fig:other_structures}f). This explains the emergence of lobs of opposite sign facing each other, which surround the central region that carries a vanishing signal due to the perfect cancellation of chiral contributions.  In the antiskyrmion case, one notices four lobs: two of positive sign  merging atop the skyrmion core, which separate the two lobs carrying negative chiral contrast. Here the spin-chirality along the x-axis passing through the core of the skyrmion is of opposite sign than that along the y-axis. 
Surprisingly,  the symmetry characterizing the antiskyrmion is not transmitted to the C-XMR signal:  The upper (left) half part is different from the lower (right) one. This originates from the electronic contribution of the Pd overlayer, where the atoms sit on a triangular lattice following the ABC stacking characterizing the assumed fcc(111) surface.

\section*{Discussion}

The chirality or topological nature of spin-textures plays a crucial role in determining their physical properties as well as their potential practical use.
Those aspects are difficult to address with current readout technologies based on the current perpendicular-to-plane geometries. Here, we discovered a rich family of spin-mixing magnetoresistances that enables an all-electrical reading of non-collinear magnetic states. Based on systematic first-principles simulations and multiple-scattering based concepts, we demonstrated the existence of a highly efficient chiral spin-mixing magnetoresistance, linear with SOC and directly induced by the chiral nature of the underlying spin-texture. We uncovered the contributions of terms quadratic with SOC that give rise to the conventional local anisotropic magnetoresistance and the newly found non-local anisotropic magnetoresistance. 
The latter is found to be orders of magnitude larger than the conventional anisotropic magnetoresistance and competes directly with the chiral or SOC-independent spin-mixing magnetoresistance in shaping the final electrical signal. The patterns emerging from each of the unveiled magnetoresistance terms as well as their distinct dependence on the misalignment of the magnetic moments can be utilized  to identify their respective amplitude and probably explain previously reported  intriguing  behaviors~\cite{Hanneken2015}. This enforces our view that such  reading mechanisms represent a new avenue for further fundamental studies of the very rich physics of chiral and topological magnetism. We envision the use of the new magnetoresistances in readily available all-electrical readout device architectures to sense, distinguish and categorize the ever increasing number of predicted and discovered non-collinear spin-textures.

\begin{methods}

\subsection{Computational details.}
The simulations are based on density functional theory (DFT) in the local spin density approximation as implemented in the full-potential  scalar-relativistic Korringa-Kohn-Rostoker (KKR) Green function method with spin-orbit coupling included self-consistently\cite{Papanikolaou2002,Bauer2013}.
The method enables the embedding of single magnetic skyrmions in a real-space approach that does not require the use of periodic supercells. The approach is based on two steps: First, self-consistent calculations of the skyrmion-free ferromagnetic slab with periodic boundary conditions are performed. The associated Green function, $G_0$, is extracted and utilized to solve the Dyson equation, schematically written as $G = G_0 + G_0 \Delta V G$, in order to obtain the new Green function $G$. $\Delta V$ represents the potential changed after injecting the magnetic skyrmion. The magnetic textures are obtained in a self-consistent fashion till convergence is achieved.

The PdFe bilayer on Ir(111) surface is simulated with a slab consisting of a layer of Pd atop Fe, which is deposited on 34 layers of Ir. We assumed an fcc-stacking for all layers  with atomic positions obtained from ab-initio~\cite{Dupe2014}. The embedded cluster consists of a total of 165 atoms, including 37 Fe atoms  (see Refs.~\cite{Crum2015,LimaFernandes2018,Fernandes2020,Fernandes2020a,Arjana2020,Bouhassoune2021} and references therein for details). We assume an angular momentum cutoff at $l_{\text{max}} = 3$ for the orbital expansion of the Green function and when extracting the local density of states a k-mesh of $200 \times 200$ is considered.

\subsection{Data availability} The data that support the findings of this study are available from the corresponding authors on request.

\subsection{Code availability} The KKR Green function code that supports the findings of this study is available from the corresponding author on request.

\end{methods}

\begin{addendum}

\item This work is supported by the European Research Council (ERC) under the European Union’s Horizon 2020 research and innovation program (ERC-consolidator grant 681405— DYNASORE and Grant No. 856538, project “3D MAGiC”) and  the Deutsche For\-schungs\-gemeinschaft (DFG) through SPP 2137 ``Skyrmionics'' (Project LO 1659/8-1). We acknowledge the computing time granted by the JARA-HPC Vergabegremium and VSR commission on the supercomputers JURECA at Forschungszentrum Jülich and at the supercomputing centre of RWTH Aachen University.

\item[Author contributions] S.L. initiated, designed and supervised the project. I.L.F. performed the first-principles calculations and post-processed the data. S.L. provided the analytical derivations of the various magnetoresistance effects. I.L.F., S.B. and S.L. discussed the results and helped writing the manuscript.

\item[Competing Interests] The authors declare no competing interests.

\item[Correspondence] Correspondence and requests for materials should be addressed to I.L.F. (email: i.lima.fernandes@fz-juelich.de) or to S.L. (email: s.lounis@fz-juelich.de).

\end{addendum}

\newpage

\section*{References}
\bibliographystyle{naturemag}

\begin{thebibliography}{10}
\expandafter\ifx\csname url\endcsname\relax
  \def\url#1{\texttt{#1}}\fi
\expandafter\ifx\csname urlprefix\endcsname\relax\def\urlprefix{URL }\fi
\providecommand{\bibinfo}[2]{#2}
\providecommand{\eprint}[2][]{\url{#2}}

\bibitem{Fert2013}
\bibinfo{author}{Fert, A.}, \bibinfo{author}{Cros, V.} \&
  \bibinfo{author}{Sampaio, J.}
\newblock \bibinfo{title}{{Skyrmions on the track}}.
\newblock \emph{\bibinfo{journal}{Nat. Nanotech.}}
  \textbf{\bibinfo{volume}{8}}, \bibinfo{pages}{152--156}
  (\bibinfo{year}{2013}).

\bibitem{Fert2017}
\bibinfo{author}{Fert, A.}, \bibinfo{author}{Reyren, N.} \&
  \bibinfo{author}{Cros, V.}
\newblock \bibinfo{title}{{Magnetic skyrmions: advances in physics and
  potential applications}}.
\newblock \emph{\bibinfo{journal}{Nat. Rev. Mater.}}
  \textbf{\bibinfo{volume}{2}}, \bibinfo{pages}{17031} (\bibinfo{year}{2017}).

\bibitem{Bogdanov1989}
\bibinfo{author}{Bogdanov, A.~N.} \& \bibinfo{author}{Yablonskii, D.}
\newblock \bibinfo{title}{Thermodynamically stable ``vortices" in magnetically
  ordered crystals. the mixed state of magnets}.
\newblock \emph{\bibinfo{journal}{J. Exp. Theor. Phys.}}
  \textbf{\bibinfo{volume}{95}}, \bibinfo{pages}{178} (\bibinfo{year}{1989}).

\bibitem{Roessler2006}
\bibinfo{author}{R\"ossler, U.~K.}, \bibinfo{author}{Bogdanov, A.~N.} \&
  \bibinfo{author}{Pfleiderer, C.}
\newblock \bibinfo{title}{Spontaneous skyrmion ground states in magnetic
  metals}.
\newblock \emph{\bibinfo{journal}{Nature}} \textbf{\bibinfo{volume}{442}},
  \bibinfo{pages}{797--801} (\bibinfo{year}{2006}).

\bibitem{Nagaosa2013}
\bibinfo{author}{Nagaosa, N.} \& \bibinfo{author}{Tokura, Y.}
\newblock \bibinfo{title}{{Topological properties and dynamics of magnetic
  skyrmions.}}
\newblock \emph{\bibinfo{journal}{Nat. Nanotech.}}
  \textbf{\bibinfo{volume}{8}}, \bibinfo{pages}{899--911}
  (\bibinfo{year}{2013}).

\bibitem{Jonietz2010}
\bibinfo{author}{Jonietz, F.} \emph{et~al.}
\newblock \bibinfo{title}{{Spin transfer torques in MnSi at ultralow current
  densities}}.
\newblock \emph{\bibinfo{journal}{Science}} \textbf{\bibinfo{volume}{330}},
  \bibinfo{pages}{1648--1651} (\bibinfo{year}{2010}).

\bibitem{Sampaio2013}
\bibinfo{author}{Sampaio, J.}, \bibinfo{author}{Cros, V.},
  \bibinfo{author}{Rohart, S.}, \bibinfo{author}{Thiaville, A.} \&
  \bibinfo{author}{Fert, A.}
\newblock \bibinfo{title}{Nucleation, stability and current-induced motion of
  isolated magnetic skyrmions in nanostructures}.
\newblock \emph{\bibinfo{journal}{Nat. Nanotech.}}
  \textbf{\bibinfo{volume}{8}}, \bibinfo{pages}{839--844}
  (\bibinfo{year}{2013}).

\bibitem{Woo2016}
\bibinfo{author}{Woo, S.} \emph{et~al.}
\newblock \bibinfo{title}{{Observation of room-temperature magnetic skyrmions
  and their current-driven dynamics in ultrathin metallic ferromagnets}}.
\newblock \emph{\bibinfo{journal}{Nat. Mater.}} \textbf{\bibinfo{volume}{15}},
  \bibinfo{pages}{501--506} (\bibinfo{year}{2016}).

\bibitem{Jiang2016}
\bibinfo{author}{Jiang, W.} \emph{et~al.}
\newblock \bibinfo{title}{{Direct observation of the skyrmion Hall effect}}.
\newblock \emph{\bibinfo{journal}{Nat. Phys.}} \textbf{\bibinfo{volume}{13}},
  \bibinfo{pages}{162--169} (\bibinfo{year}{2017}).

\bibitem{Litzius2017}
\bibinfo{author}{Litzius, K.} \emph{et~al.}
\newblock \bibinfo{title}{Skyrmion hall effect revealed by direct time-resolved
  x-ray microscopy}.
\newblock \emph{\bibinfo{journal}{Nat. Phys.}} \textbf{\bibinfo{volume}{13}},
  \bibinfo{pages}{170--175} (\bibinfo{year}{2017}).

\bibitem{Ma2018}
\bibinfo{author}{Ma, C.} \emph{et~al.}
\newblock \bibinfo{title}{Electric field-induced creation and directional
  motion of domain walls and skyrmion bubbles}.
\newblock \emph{\bibinfo{journal}{Nano letters}} \textbf{\bibinfo{volume}{19}},
  \bibinfo{pages}{353--361} (\bibinfo{year}{2018}).

\bibitem{Hu2018}
\bibinfo{author}{Hu, J.-M.}, \bibinfo{author}{Yang, T.} \&
  \bibinfo{author}{Chen, L.-Q.}
\newblock \bibinfo{title}{Strain-mediated voltage-controlled switching of
  magnetic skyrmions in nanostructures}.
\newblock \emph{\bibinfo{journal}{npj Computational Materials}}
  \textbf{\bibinfo{volume}{4}}, \bibinfo{pages}{1--7} (\bibinfo{year}{2018}).

\bibitem{Koshibae2014}
\bibinfo{author}{Koshibae, W.} \& \bibinfo{author}{Nagaosa, N.}
\newblock \bibinfo{title}{Creation of skyrmions and antiskyrmions by local
  heating}.
\newblock \emph{\bibinfo{journal}{Nature Communications}}
  \textbf{\bibinfo{volume}{5}}, \bibinfo{pages}{5148} (\bibinfo{year}{2014}).
\newblock \urlprefix\url{https://doi.org/10.1038/ncomms6148}.

\bibitem{Leonov2015}
\bibinfo{author}{Leonov, A.~O.} \& \bibinfo{author}{Mostovoy, M.}
\newblock \bibinfo{title}{Multiply periodic states and isolated skyrmions in an
  anisotropic frustrated magnet}.
\newblock \emph{\bibinfo{journal}{Nature Communications}}
  \textbf{\bibinfo{volume}{6}}, \bibinfo{pages}{8275} (\bibinfo{year}{2015}).
\newblock \urlprefix\url{https://doi.org/10.1038/ncomms9275}.

\bibitem{Ritzmann2018}
\bibinfo{author}{Ritzmann, U.} \emph{et~al.}
\newblock \bibinfo{title}{Trochoidal motion and pair generation in skyrmion and
  antiskyrmion dynamics under spin--orbit torques}.
\newblock \emph{\bibinfo{journal}{Nature Electronics}}
  \textbf{\bibinfo{volume}{1}}, \bibinfo{pages}{451--457}
  (\bibinfo{year}{2018}).
\newblock \urlprefix\url{https://doi.org/10.1038/s41928-018-0114-0}.

\bibitem{Rybakov2019}
\bibinfo{author}{Rybakov, F.~N.} \& \bibinfo{author}{Kiselev, N.~S.}
\newblock \bibinfo{title}{Chiral magnetic skyrmions with arbitrary topological
  charge}.
\newblock \emph{\bibinfo{journal}{Phys. Rev. B}} \textbf{\bibinfo{volume}{99}},
  \bibinfo{pages}{064437} (\bibinfo{year}{2019}).
\newblock \urlprefix\url{https://link.aps.org/doi/10.1103/PhysRevB.99.064437}.

\bibitem{Yokouchi2020}
\bibinfo{author}{Yokouchi, T.} \emph{et~al.}
\newblock \bibinfo{title}{Creation of magnetic skyrmions by surface acoustic
  waves}.
\newblock \emph{\bibinfo{journal}{Nature Nanotechnology}}
  \textbf{\bibinfo{volume}{15}}, \bibinfo{pages}{361--366}
  (\bibinfo{year}{2020}).
\newblock \urlprefix\url{https://doi.org/10.1038/s41565-020-0661-1}.

\bibitem{Ritzmann2020}
\bibinfo{author}{Ritzmann, U.}, \bibinfo{author}{Desplat, L.},
  \bibinfo{author}{Dup\'e, B.}, \bibinfo{author}{Camley, R.~E.} \&
  \bibinfo{author}{Kim, J.-V.}
\newblock \bibinfo{title}{Asymmetric skyrmion-antiskyrmion production in
  ultrathin ferromagnetic films}.
\newblock \emph{\bibinfo{journal}{Phys. Rev. B}}
  \textbf{\bibinfo{volume}{102}}, \bibinfo{pages}{174409}
  (\bibinfo{year}{2020}).
\newblock \urlprefix\url{https://link.aps.org/doi/10.1103/PhysRevB.102.174409}.

\bibitem{jena2020}
\bibinfo{author}{Jena, J.} \emph{et~al.}
\newblock \bibinfo{title}{Elliptical bloch skyrmion chiral twins in an
  antiskyrmion system}.
\newblock \emph{\bibinfo{journal}{Nat. Commun.}} \textbf{\bibinfo{volume}{11}},
  \bibinfo{pages}{1--9} (\bibinfo{year}{2020}).

\bibitem{Gao2020}
\bibinfo{author}{Gao, S.} \emph{et~al.}
\newblock \bibinfo{title}{Fractional antiferromagnetic skyrmion lattice induced
  by anisotropic couplings}.
\newblock \emph{\bibinfo{journal}{Nature}} \textbf{\bibinfo{volume}{586}},
  \bibinfo{pages}{37--41} (\bibinfo{year}{2020}).
\newblock \urlprefix\url{https://doi.org/10.1038/s41586-020-2716-8}.

\bibitem{Heigl2021}
\bibinfo{author}{{Heigl, Michael and Koraltan, Sabri and Va{\v{n}}atka, Marek
  and Kraft, Robert and Abert, Claas and Vogler, Christoph and Semisalova, Anna
  and Che, Ping and Ullrich, Aladin and Schmidt, Timo and others}}.
\newblock \bibinfo{title}{Dipolar-stabilized first and second-order
  antiskyrmions in ferrimagnetic multilayers}.
\newblock \emph{\bibinfo{journal}{Nat. Commun.}} \textbf{\bibinfo{volume}{12}},
  \bibinfo{pages}{1--9} (\bibinfo{year}{2021}).

\bibitem{Yao2020}
\bibinfo{author}{Yao, X.}, \bibinfo{author}{Chen, J.} \& \bibinfo{author}{Dong,
  S.}
\newblock \bibinfo{title}{Controlling the helicity of magnetic skyrmions by
  electrical field in frustrated magnets}.
\newblock \emph{\bibinfo{journal}{New Journal of Physics}}
  \textbf{\bibinfo{volume}{22}}, \bibinfo{pages}{083032}
  (\bibinfo{year}{2020}).

\bibitem{Dzyaloshinsky1958}
\bibinfo{author}{Dzyaloshinsky, I.}
\newblock \bibinfo{title}{A thermodynamic theory of “weak” ferromagnetism
  of antiferromagnetics}.
\newblock \emph{\bibinfo{journal}{J. Phys. Chem. Sol.}}
  \textbf{\bibinfo{volume}{4}}, \bibinfo{pages}{241 -- 255}
  (\bibinfo{year}{1958}).

\bibitem{Moriya1960}
\bibinfo{author}{Moriya, T.}
\newblock \bibinfo{title}{Anisotropic superexchange interaction and weak
  ferromagnetism}.
\newblock \emph{\bibinfo{journal}{Phys. Rev.}} \textbf{\bibinfo{volume}{120}},
  \bibinfo{pages}{91--98} (\bibinfo{year}{1960}).

\bibitem{Chen2021}
\bibinfo{author}{Chen, G.} \emph{et~al.}
\newblock \bibinfo{title}{Observation of hydrogen-induced dzyaloshinskii-moriya
  interaction and reversible switching of magnetic chirality}.
\newblock \emph{\bibinfo{journal}{Physical Review X}}
  \textbf{\bibinfo{volume}{11}}, \bibinfo{pages}{021015}
  (\bibinfo{year}{2021}).

\bibitem{Srivastava2018}
\bibinfo{author}{Srivastava, T.} \emph{et~al.}
\newblock \bibinfo{title}{Large-voltage tuning of dzyaloshinskii--moriya
  interactions: A route toward dynamic control of skyrmion chirality}.
\newblock \emph{\bibinfo{journal}{Nano Letters}} \textbf{\bibinfo{volume}{18}},
  \bibinfo{pages}{4871--4877} (\bibinfo{year}{2018}).
\newblock \urlprefix\url{https://doi.org/10.1021/acs.nanolett.8b01502}.

\bibitem{Buettner2021}
\bibinfo{author}{B{\"u}ttner, F.} \emph{et~al.}
\newblock \bibinfo{title}{Observation of fluctuation-mediated picosecond
  nucleation of a topological phase}.
\newblock \emph{\bibinfo{journal}{Nature Materials}}
  \textbf{\bibinfo{volume}{20}}, \bibinfo{pages}{30--37}
  (\bibinfo{year}{2021}).
\newblock \urlprefix\url{https://doi.org/10.1038/s41563-020-00807-1}.

\bibitem{Chshiev2018}
\bibinfo{author}{Yang, H.}, \bibinfo{author}{Boulle, O.},
  \bibinfo{author}{Cros, V.}, \bibinfo{author}{Fert, A.} \&
  \bibinfo{author}{Chshiev, M.}
\newblock \bibinfo{title}{Controlling dzyaloshinskii-moriya interaction via
  chirality dependent atomic-layer stacking, insulator capping and electric
  field}.
\newblock \emph{\bibinfo{journal}{Scientific reports}}
  \textbf{\bibinfo{volume}{8}}, \bibinfo{pages}{1--7} (\bibinfo{year}{2018}).

\bibitem{Desplat2020}
\bibinfo{author}{Desplat, L.} \emph{et~al.}
\newblock \bibinfo{title}{Dzyaloshinskii-moriya interaction induced by an
  ultrashort electromagnetic pulse: Application to coherent (anti)ferromagnetic
  skyrmion nucleation}.
\newblock \emph{\bibinfo{journal}{arXiv preprint arXiv:2011.12055}}
  (\bibinfo{year}{2020}).

\bibitem{Heinze2011}
\bibinfo{author}{Heinze, S.} \emph{et~al.}
\newblock \bibinfo{title}{Spontaneous atomic-scale magnetic skyrmion lattice in
  two dimensions}.
\newblock \emph{\bibinfo{journal}{Nat. Phys.}} \textbf{\bibinfo{volume}{7}},
  \bibinfo{pages}{713--718} (\bibinfo{year}{2011}).

\bibitem{Romming2013}
\bibinfo{author}{Romming, N.} \emph{et~al.}
\newblock \bibinfo{title}{Writing and deleting single magnetic skyrmions}.
\newblock \emph{\bibinfo{journal}{Science}} \textbf{\bibinfo{volume}{341}},
  \bibinfo{pages}{636--639} (\bibinfo{year}{2013}).

\bibitem{Romming2015}
\bibinfo{author}{Romming, N.}, \bibinfo{author}{Kubetzka, A.},
  \bibinfo{author}{Hanneken, C.}, \bibinfo{author}{von Bergmann, K.} \&
  \bibinfo{author}{Wiesendanger, R.}
\newblock \bibinfo{title}{Field-dependent size and shape of single magnetic
  skyrmions}.
\newblock \emph{\bibinfo{journal}{Phys. Rev. Lett.}}
  \textbf{\bibinfo{volume}{114}}, \bibinfo{pages}{177203}
  (\bibinfo{year}{2015}).

\bibitem{Meyer2019}
\bibinfo{author}{Meyer, S.} \emph{et~al.}
\newblock \bibinfo{title}{Isolated zero field sub-10 nm skyrmions in ultrathin
  co films}.
\newblock \emph{\bibinfo{journal}{Nature Communications}}
  \textbf{\bibinfo{volume}{10}}, \bibinfo{pages}{3823} (\bibinfo{year}{2019}).
\newblock \urlprefix\url{https://doi.org/10.1038/s41467-019-11831-4}.

\bibitem{Perini2019}
\bibinfo{author}{Perini, M.} \emph{et~al.}
\newblock \bibinfo{title}{Electrical detection of domain walls and skyrmions in
  co films using noncollinear magnetoresistance}.
\newblock \emph{\bibinfo{journal}{Phys. Rev. Lett.}}
  \textbf{\bibinfo{volume}{123}}, \bibinfo{pages}{237205}
  (\bibinfo{year}{2019}).
\newblock
  \urlprefix\url{https://link.aps.org/doi/10.1103/PhysRevLett.123.237205}.

\bibitem{Muckel2021}
\bibinfo{author}{Muckel, F.} \emph{et~al.}
\newblock \bibinfo{title}{Experimental identification of two distinct skyrmion
  collapse mechanisms}.
\newblock \emph{\bibinfo{journal}{Nature Physics}}
  \textbf{\bibinfo{volume}{17}}, \bibinfo{pages}{395--402}
  (\bibinfo{year}{2021}).
\newblock \urlprefix\url{https://doi.org/10.1038/s41567-020-01101-2}.

\bibitem{McGuire1975}
\bibinfo{author}{McGuire, T.} \& \bibinfo{author}{Potter, R.}
\newblock \bibinfo{title}{Anisotropic magnetoresistance in ferromagnetic 3d
  alloys}.
\newblock \emph{\bibinfo{journal}{IEEE Transactions on Magnetics}}
  \textbf{\bibinfo{volume}{11}}, \bibinfo{pages}{1018--1038}
  (\bibinfo{year}{1975}).

\bibitem{Bode2002}
\bibinfo{author}{Bode, M.} \emph{et~al.}
\newblock \bibinfo{title}{Magnetization-direction-dependent local electronic
  structure probed by scanning tunneling spectroscopy}.
\newblock \emph{\bibinfo{journal}{Phys. Rev. Lett.}}
  \textbf{\bibinfo{volume}{89}}, \bibinfo{pages}{237205}
  (\bibinfo{year}{2002}).
\newblock
  \urlprefix\url{https://link.aps.org/doi/10.1103/PhysRevLett.89.237205}.

\bibitem{Gould2004}
\bibinfo{author}{Gould, C.} \emph{et~al.}
\newblock \bibinfo{title}{Tunneling anisotropic magnetoresistance: A
  spin-valve-like tunnel magnetoresistance using a single magnetic layer}.
\newblock \emph{\bibinfo{journal}{Phys. Rev. Lett.}}
  \textbf{\bibinfo{volume}{93}}, \bibinfo{pages}{117203}
  (\bibinfo{year}{2004}).
\newblock
  \urlprefix\url{https://link.aps.org/doi/10.1103/PhysRevLett.93.117203}.

\bibitem{Matos-Abiague2009}
\bibinfo{author}{Matos-Abiague, A.} \& \bibinfo{author}{Fabian, J.}
\newblock \bibinfo{title}{Anisotropic tunneling magnetoresistance and tunneling
  anisotropic magnetoresistance: Spin-orbit coupling in magnetic tunnel
  junctions}.
\newblock \emph{\bibinfo{journal}{Phys. Rev. B}} \textbf{\bibinfo{volume}{79}},
  \bibinfo{pages}{155303} (\bibinfo{year}{2009}).
\newblock \urlprefix\url{https://link.aps.org/doi/10.1103/PhysRevB.79.155303}.

\bibitem{Crum2015}
\bibinfo{author}{Crum, D.~M.} \emph{et~al.}
\newblock \bibinfo{title}{{Perpendicular reading of single confined magnetic
  skyrmions}}.
\newblock \emph{\bibinfo{journal}{Nat. Commun.}} \textbf{\bibinfo{volume}{6}},
  \bibinfo{pages}{8541} (\bibinfo{year}{2015}).

\bibitem{Hanneken2015}
\bibinfo{author}{Hanneken, C.} \emph{et~al.}
\newblock \bibinfo{title}{Electrical detection of magnetic skyrmions by
  tunnelling non-collinear magnetoresistance}.
\newblock \emph{\bibinfo{journal}{Nat. Nanotechnol.}}
  \textbf{\bibinfo{volume}{10}}, \bibinfo{pages}{1039} (\bibinfo{year}{2015}).

\bibitem{Kubetzka2017}
\bibinfo{author}{Kubetzka, A.}, \bibinfo{author}{Hanneken, C.},
  \bibinfo{author}{Wiesendanger, R.} \& \bibinfo{author}{von Bergmann, K.}
\newblock \bibinfo{title}{Impact of the skyrmion spin texture on
  magnetoresistance}.
\newblock \emph{\bibinfo{journal}{Phys. Rev. B}} \textbf{\bibinfo{volume}{95}},
  \bibinfo{pages}{104433} (\bibinfo{year}{2017}).

\bibitem{Bruno2004}
\bibinfo{author}{Bruno, P.}, \bibinfo{author}{Dugaev, V.~K.} \&
  \bibinfo{author}{Taillefumier, M.}
\newblock \bibinfo{title}{Topological hall effect and berry phase in magnetic
  nanostructures}.
\newblock \emph{\bibinfo{journal}{Phys. Rev. Lett.}}
  \textbf{\bibinfo{volume}{93}}, \bibinfo{pages}{096806}
  (\bibinfo{year}{2004}).
\newblock
  \urlprefix\url{https://link.aps.org/doi/10.1103/PhysRevLett.93.096806}.

\bibitem{Lux2020}
\bibinfo{author}{Lux, F.~R.}, \bibinfo{author}{Freimuth, F.},
  \bibinfo{author}{Bl\"ugel, S.} \& \bibinfo{author}{Mokrousov, Y.}
\newblock \bibinfo{title}{Chiral hall effect in noncollinear magnets from a
  cyclic cohomology approach}.
\newblock \emph{\bibinfo{journal}{Phys. Rev. Lett.}}
  \textbf{\bibinfo{volume}{124}}, \bibinfo{pages}{096602}
  (\bibinfo{year}{2020}).
\newblock
  \urlprefix\url{https://link.aps.org/doi/10.1103/PhysRevLett.124.096602}.

\bibitem{Bouaziz2021}
\bibinfo{author}{Bouaziz, J.}, \bibinfo{author}{Ishida, H.},
  \bibinfo{author}{Lounis, S.} \& \bibinfo{author}{Bl\"ugel, S.}
\newblock \bibinfo{title}{Transverse transport in two-dimensional relativistic
  systems with nontrivial spin textures}.
\newblock \emph{\bibinfo{journal}{Phys. Rev. Lett.}}
  \textbf{\bibinfo{volume}{126}}, \bibinfo{pages}{147203}
  (\bibinfo{year}{2021}).
\newblock
  \urlprefix\url{https://link.aps.org/doi/10.1103/PhysRevLett.126.147203}.

\bibitem{Neubauer2009}
\bibinfo{author}{Neubauer, A.} \emph{et~al.}
\newblock \bibinfo{title}{Topological hall effect in the $a$ phase of mnsi}.
\newblock \emph{\bibinfo{journal}{Phys. Rev. Lett.}}
  \textbf{\bibinfo{volume}{102}}, \bibinfo{pages}{186602}
  (\bibinfo{year}{2009}).
\newblock
  \urlprefix\url{https://link.aps.org/doi/10.1103/PhysRevLett.102.186602}.

\bibitem{Lee2009}
\bibinfo{author}{Lee, M.}, \bibinfo{author}{Kang, W.}, \bibinfo{author}{Onose,
  Y.}, \bibinfo{author}{Tokura, Y.} \& \bibinfo{author}{Ong, N.~P.}
\newblock \bibinfo{title}{Unusual hall effect anomaly in mnsi under pressure}.
\newblock \emph{\bibinfo{journal}{Phys. Rev. Lett.}}
  \textbf{\bibinfo{volume}{102}}, \bibinfo{pages}{186601}
  (\bibinfo{year}{2009}).
\newblock
  \urlprefix\url{https://link.aps.org/doi/10.1103/PhysRevLett.102.186601}.

\bibitem{Zhang2016}
\bibinfo{author}{Zhang, S.~F.} \emph{et~al.}
\newblock \bibinfo{title}{Highly efficient domain walls injection in
  perpendicular magnetic anisotropy nanowire}.
\newblock \emph{\bibinfo{journal}{Scientific Reports}}
  \textbf{\bibinfo{volume}{6}}, \bibinfo{pages}{24804} (\bibinfo{year}{2016}).
\newblock \urlprefix\url{https://doi.org/10.1038/srep24804}.

\bibitem{Maccariello2018}
\bibinfo{author}{Maccariello, D.} \emph{et~al.}
\newblock \bibinfo{title}{Electrical detection of single magnetic skyrmions in
  metallic multilayers at room temperature}.
\newblock \emph{\bibinfo{journal}{Nat. Nanotechnol.}}
  \textbf{\bibinfo{volume}{13}}, \bibinfo{pages}{233} (\bibinfo{year}{2018}).

\bibitem{Zeissler2018}
\bibinfo{author}{Zeissler, K.} \emph{et~al.}
\newblock \bibinfo{title}{Discrete hall resistivity contribution from n{\'e}el
  skyrmions in multilayer nanodiscs}.
\newblock \emph{\bibinfo{journal}{Nature Nanotechnology}}
  \textbf{\bibinfo{volume}{13}}, \bibinfo{pages}{1161--1166}
  (\bibinfo{year}{2018}).
\newblock \urlprefix\url{https://doi.org/10.1038/s41565-018-0268-y}.

\bibitem{Dupe2014}
\bibinfo{author}{Dup{\'e}, B.}, \bibinfo{author}{Hoffmann, M.},
  \bibinfo{author}{Paillard, C.} \& \bibinfo{author}{Heinze, S.}
\newblock \bibinfo{title}{Tailoring magnetic skyrmions in ultra-thin transition
  metal films}.
\newblock \emph{\bibinfo{journal}{Nat. Commun.}} \textbf{\bibinfo{volume}{5}},
  \bibinfo{pages}{4030} (\bibinfo{year}{2014}).

\bibitem{Simon2014}
\bibinfo{author}{Simon, E.}, \bibinfo{author}{Palot{\'a}s, K.},
  \bibinfo{author}{R{\'o}zsa, L.}, \bibinfo{author}{Udvardi, L.} \&
  \bibinfo{author}{Szunyogh, L.}
\newblock \bibinfo{title}{{ Formation of magnetic skyrmions with tunable
  properties in PdFe bilayer deposited on Ir (111)}}.
\newblock \emph{\bibinfo{journal}{Phys. Rev. B}} \textbf{\bibinfo{volume}{90}},
  \bibinfo{pages}{094410} (\bibinfo{year}{2014}).

\bibitem{Dias2016}
\bibinfo{author}{dos Santos~Dias, M.}, \bibinfo{author}{Bouaziz, J.},
  \bibinfo{author}{Bouhassoune, M.}, \bibinfo{author}{Bl\"ugel, S.} \&
  \bibinfo{author}{Lounis, S.}
\newblock \bibinfo{title}{Chirality-driven orbital magnetic moments as a new
  probe for topological magnetic structures}.
\newblock \emph{\bibinfo{journal}{Nature Commun.}}
  \textbf{\bibinfo{volume}{7}}, \bibinfo{pages}{13613} (\bibinfo{year}{2016}).

\bibitem{Tersoff1983}
\bibinfo{author}{Tersoff, J.} \& \bibinfo{author}{Hamann, D.}
\newblock \bibinfo{title}{Theory and application for the scanning tunneling
  microscope}.
\newblock \emph{\bibinfo{journal}{Phys. Rev. Lett.}}
  \textbf{\bibinfo{volume}{50}}, \bibinfo{pages}{1998} (\bibinfo{year}{1983}).

\bibitem{Papanikolaou2002}
\bibinfo{author}{Papanikolaou, N.}, \bibinfo{author}{Zeller, R.} \&
  \bibinfo{author}{Dederichs, P.~H.}
\newblock \bibinfo{title}{{Conceptual improvements of the KKR method}}.
\newblock \emph{\bibinfo{journal}{J. Phys. Condens. Matter}}
  \textbf{\bibinfo{volume}{14}}, \bibinfo{pages}{2799} (\bibinfo{year}{2002}).

\bibitem{Bauer2013}
\bibinfo{author}{Bauer, D. S.~G.}
\newblock \bibinfo{title}{Development of a relativistic full-potential
  first-principles multiple scattering green function method applied to complex
  magnetic textures of nano structures at surfaces}.
\newblock \emph{\bibinfo{journal}{PhD dissertation at the RWTH-Aachen}}
  (\bibinfo{year}{2013}).

\bibitem{LimaFernandes2018}
\bibinfo{author}{{Lima Fernandes}, I.}, \bibinfo{author}{Bouaziz, J.},
  \bibinfo{author}{Bl{\"{u}}gel, S.} \& \bibinfo{author}{Lounis, S.}
\newblock \bibinfo{title}{{Universality of defect-skyrmion interaction
  profiles}}.
\newblock \emph{\bibinfo{journal}{Nat. Commun.}} \textbf{\bibinfo{volume}{9}},
  \bibinfo{pages}{4395} (\bibinfo{year}{2018}).

\bibitem{Fernandes2020}
\bibinfo{author}{Fernandes, I.~L.}, \bibinfo{author}{Bouhassoune, M.} \&
  \bibinfo{author}{Lounis, S.}
\newblock \bibinfo{title}{Defect-implantation for the all-electrical detection
  of non-collinear spin-textures}.
\newblock \emph{\bibinfo{journal}{Nature Commun.}}
  \textbf{\bibinfo{volume}{11}}, \bibinfo{pages}{1--9} (\bibinfo{year}{2020}).

\bibitem{Fernandes2020a}
\bibinfo{author}{Fernandes, I.~L.}, \bibinfo{author}{Chico, J.} \&
  \bibinfo{author}{Lounis, S.}
\newblock \bibinfo{title}{{Impurity-dependent gyrotropic motion, deflection and
  pinning of current-driven ultrasmall skyrmions in PdFe/Ir(111) surface}}.
\newblock \emph{\bibinfo{journal}{Journal of Physics: Condensed Matter}}
  (\bibinfo{year}{2020}).
\newblock \urlprefix\url{http://iopscience.iop.org/10.1088/1361-648X/ab9cf0}.

\bibitem{Arjana2020}
\bibinfo{author}{Arjana, I.~G.}, \bibinfo{author}{Lima~Fernandes, I.},
  \bibinfo{author}{Chico, J.} \& \bibinfo{author}{Lounis, S.}
\newblock \bibinfo{title}{Sub-nanoscale atom-by-atom crafting of
  skyrmion-defect interaction profiles}.
\newblock \emph{\bibinfo{journal}{Scientific Reports}}
  \textbf{\bibinfo{volume}{10}}, \bibinfo{pages}{14655} (\bibinfo{year}{2020}).
\newblock \urlprefix\url{https://doi.org/10.1038/s41598-020-71232-2}.

\bibitem{Bouhassoune2021}
\bibinfo{author}{Bouhassoune, M.} \& \bibinfo{author}{Lounis, S.}
\newblock \bibinfo{title}{Friedel oscillations induced by magnetic skyrmions:
  From scattering properties to all-electrical detection}.
\newblock \emph{\bibinfo{journal}{Nanomaterials}} \textbf{\bibinfo{volume}{11}}
  (\bibinfo{year}{2021}).
\newblock \urlprefix\url{https://www.mdpi.com/2079-4991/11/1/194}.

\end{thebibliography}

\end{document}